\documentclass[conference]{IEEEtran}
\IEEEoverridecommandlockouts
\usepackage{cite}
\usepackage{amsmath,amssymb,amsfonts}
\usepackage{algorithmic}
\usepackage{graphicx}
\usepackage{textcomp}
\usepackage{xcolor}
\usepackage{hyperref}

\def\BibTeX{{\rm B\kern-.05em{\sc i\kern-.025em b}\kern-.08em
    T\kern-.1667em\lower.7ex\hbox{E}\kern-.125emX}}
\begin{document}

\title{
    DataDock: An Open Source Data Hub for Research
}

\author{\IEEEauthorblockN{Lexington Whalen}
\IEEEauthorblockA{\textit{Department of Computer Science} \\
\textit{University of South Carolina}\\
Columbia, United States\\
LAWHALEN@email.sc.edu}
\and
\IEEEauthorblockN{Homayoun Valafar}
\IEEEauthorblockA{\textit{Department of Computer Science} \\
\textit{University of South Carolina}\\
Columbia, United States\\
homayoun@cse.sc.edu}
}

\maketitle

\begin{abstract}
Every research project necessitates data, often requiring sharing and collaborative review within a team. 
However, there is a dearth of good open-source data sharing and reviewing services. Existing file-sharing services generally mandate paid subscriptions for increased storage or additional members, 
diverting research funds from addressing the core research problem that a lab is attempting to work on. 
Moreover, these services often lack direct features for reviewing or 
commenting on data quality, a vital part of ensuring high quality data generation. 
In response to these challenges, we present DataDock, a specialized file transfer service crafted for specifically for researchers. 
DataDock operates as an application hosted on a research lab server. This design ensures that, with access to a machine and an internet connection, 
teams can facilitate file storage, transfer, and review without incurring extra costs. Being an open-source project, DataDock can be customized to 
suit the unique requirements of any research team, and is able to evolve to meet the needs of the research community. We also note that there are no 
limitations with respect to what data can be shared, downloaded, or commented on. As DataDock is agnostic to the file type, it can be used in any field from 
bioinformatics to particle physics; as long as it can be stored in a file, it can be shared.
We open source the code here: 
\href{https://github.com/lxaw/DataDock}{https://github.com/lxaw/DataDock}.

\end{abstract}

\begin{IEEEkeywords}
data transfer, data storage, data reviewing, data sharing, data governance, data maintenance
\end{IEEEkeywords}

\section{Introduction}
There is a noticeable dearth of quality open source research services with regards to data.
Most data services are merely file sharing ones, that only allow file upload, download, and organization.
While this does allow researchers to transmit their collected data among another,
it lacks many critical features that most researchers would like to have, such as
quality control, commenting, tagging, organization under groups, annotation, and more.
Furthermore, these services are oftentimes prohibitively expensive to smaller research labs, and their
free versions have limitations on both number of users and amount of storage.
The current state of research is to use services such as Microsoft's OneDrive \cite{onedrive}, Dropbox \cite{dropbox},
or Google Drive \cite{googledrive}. Each of these services have limitations in terms of storage capacity, user management,
and feature set that make them suboptimal for research data management \cite{kowalczyk2018data}.
\\
Furthermore, the modern times have seen unprecedented growth in the amount of data collected.
The era of big data has brought about massive datasets, particularly in fields like genomics, astronomy, and social media analytics \cite{stephens2015big}. 
The diversity of the data being stored is also of importance. A single lab may deal with large nutrition datasets \cite{foodData}, audio datasets \cite{wordification}, EEG data \cite{eeg}, or cancer data \cite{cancer}.
Machine learning and deep learning have further amplified the need for large, high-quality datasets for training models \cite{sun2017revisiting}.
The Internet of Things (IoT) is another significant contributor to the data deluge, with billions of connected devices generating continuous streams of data \cite{khan2018iot}.
Even at a personal level, the proliferation of smartphones and digital services has led to an explosion of user-generated data \cite{reinsel2018digitization}.
\\
This rapid growth in data volume and variety has outpaced the capabilities of traditional data management tools and practices.
Researchers often struggle with the challenges of storing, organizing, sharing, and collaborating on large datasets \cite{wallis2013if}.
The lack of specialized tools for research data management leads to ad-hoc solutions, data silos, and inefficiencies in the research process \cite{wilkinson2016fair}.
\\
To address these challenges, we present DataDock, an open source data service designed specifically for the needs of researchers.
DataDock provides a one-stop platform for efficient data storage, sharing, and collaboration, with features tailored to the research workflow.
By leveraging open source technologies and a modular architecture, DataDock enables customization and extensibility to meet the diverse requirements of different research domains.
The open source nature of DataDock also promotes transparency, reproducibility, and community-driven development \cite{gezelter2015open}.

\section{The Problem}
In the era of data-driven research, scientists across various disciplines are grappling with an unprecedented influx of data. From high-throughput sequencing in genomics \cite{stephens2015big} to large-scale simulations in physics \cite{vogelsberger2014introducing}, researchers are generating and collecting massive datasets at an astounding rate. However, managing this data effectively poses significant challenges that can hinder research productivity and collaboration.

One of the primary issues researchers face is the need for efficient data labeling and annotation. Raw data often requires contextualization and metadata to be meaningful and usable for analysis \cite{wilkinson2016fair}. Manually annotating large datasets is time-consuming and prone to inconsistencies, especially when multiple researchers are involved. The lack of standardized annotation frameworks and tools further compounds this problem, leading to a fragmented and inefficient data labeling process.

Quality control is another critical concern in research data management. Ensuring the accuracy, completeness, and reliability of data is essential for drawing valid conclusions and reproducing results \cite{eisner2018data}. However, reviewing and validating large datasets manually is a daunting task, particularly in collaborative projects where data is collected by multiple individuals or teams. The absence of systematic quality control mechanisms can lead to errors, inconsistencies, and a lack of trust in the data.

Effective organization and management of research data is also a significant challenge, both at the individual researcher level and within research groups. Principal Investigators (PIs) need to keep track of various datasets, their versions, and associated metadata, often relying on ad-hoc solutions like spreadsheets or file naming conventions \cite{kowalczyk2018data}. At the group level, data silos can emerge when different teams use incompatible data formats or storage systems, hindering collaboration and data integration.

Sharing data between research groups is another pain point in the current research landscape. Collaborative projects often involve multiple institutions and disciplines, each with their own data management practices and infrastructure \cite{tenopir2011data}. Incompatible data formats, lack of standardization, and concerns about data ownership and attribution can impede smooth data sharing and reuse. Moreover, transferring large datasets across institutional boundaries can be time-consuming and subject to security and privacy constraints.

Addressing these challenges requires a comprehensive and integrated approach to research data management. Researchers need tools and platforms that streamline data annotation, enable robust quality control, facilitate organization and discovery, and promote seamless data sharing and collaboration. Existing solutions, such as generic cloud storage services or ad-hoc scripts, often fall short in terms of functionality, scalability, and ease of use. There is a clear need for a purpose-built, open-source data management platform that caters to the unique requirements of scientific research.

\section{Current Alternatives}
Several cloud storage providers offer APIs and developer tools that enable programmatic access to their services. For example:

\begin{itemize}
\item Google Drive provides a REST API \cite{googledrive} for uploading, downloading, searching, and manipulating files stored on Google Drive. Client libraries are available in multiple programming languages.

\item Dropbox offers a similar REST API \cite{dropbox} with SDKs for major platforms to integrate Dropbox capabilities into applications. Key features include file upload/download, sharing, search, and user management.

\item Microsoft OneDrive also has a comprehensive REST API \cite{onedrive} for accessing OneDrive files, folders, and other data. Client libraries support integration with web, mobile, and desktop apps.

\item Box provides a content management platform with a REST API \cite{box} for building custom applications. It offers features like file preview, version control, and granular access permissions, catering to enterprise needs.

\item Amazon S3 (Simple Storage Service) is a scalable object storage service with an API \cite{amazonweb} for storing and retrieving data. While not primarily designed for end-user file management, it's often used as a foundation for building cloud storage applications.
\end{itemize}

While these APIs enable building custom applications with cloud storage, the underlying limitations of the services still apply - costs scale with storage and user needs, and specialized features for research data management are lacking. Leveraging the APIs still requires significant development effort to create a tailored solution. Furthermore, some services like Amazon S3 are more suited to developers and lack user-friendly interfaces out-of-the-box.
Furthermore, these services are not open sourced, so any modifications (i.e. addition of service, design change, bug fix) cannot be done directly by the users.
We make such issues more clear in $\textit{Issues with Current Approaches}$.

\section{Issues with Current Approaches}
While current cloud storage services offer APIs and enable building custom applications, they have several notable shortcomings, especially in the context of research data management:

\begin{enumerate}
\item \textbf{Cost:} Services like Google Drive, Dropbox, and OneDrive can become expensive as data storage needs grow, often requiring paid subscriptions for additional storage or users. This diverts funds from core research activities.

\item \textbf{Lack of Specialization:} General-purpose storage services lack features tailored for research, such as data quality control, peer review workflows, metadata management, and data provenance tracking.

\item \textbf{Limited Customization:} Although APIs enable custom app development, the underlying platforms cannot be easily modified or extended. Researchers cannot add new features or modify existing behavior to suit their specific needs.

\item \textbf{Vendor Lock-In:} By relying on proprietary services, researchers risk being locked into a particular vendor's ecosystem. Migration to alternative platforms can be difficult and costly.

\item \textbf{Data Ownership and Control:} With commercial services, there may be ambiguity around data ownership and control. Researchers may have concerns about intellectual property rights and the ability to access their data if a service is discontinued.

\item \textbf{Data Privacy and Security:} Storing sensitive research data on third-party servers raises privacy and security concerns. Researchers may be hesitant to entrust confidential data to external providers.

\item \textbf{Collaboration Barriers:} While cloud services facilitate file sharing, they often lack advanced collaboration features like real-time co-authoring, version control, and granular access controls that are vital for research teams.

\item \textbf{Integration Challenges:} Integrating cloud storage with existing research tools and workflows can be challenging. Researchers may need to develop custom glue code or rely on limited third-party integrations.

\item \textbf{Dependency on Internet Connectivity:} Cloud services require reliable internet access, which can be a constraint in field research settings or areas with limited connectivity.

\item \textbf{Long-Term Preservation:} Commercial services may not prioritize long-term data preservation, which is crucial for research reproducibility and data archiving. There may be uncertainties about data durability and accessibility over extended periods.
\end{enumerate}

An open source solution like DataDock addresses these issues by providing a specialized, customizable, and cost-effective platform for research data management. By hosting DataDock on their own infrastructure, research teams have full control over their data, can tailor the platform to their specific needs, and avoid vendor lock-in. The open source nature ensures transparency, enables community-driven development, and allows for integration with a wide range of tools. Moreover, an open source solution can be deployed in local or offline environments, mitigating concerns about internet connectivity and data privacy. DataDock empowers researchers to manage their data on their own terms, prioritizing the unique requirements of scientific research.

\section{DataDock Architecture}
DataDock is built using a modern web stack consisting of Django \cite{django}, Django REST Framework \cite{djangorest}, React \cite{react}, and Redux \cite{redux}. This architecture was chosen to enable fast, scalable development, cross-platform compatibility, and a smooth user experience.

\subsection{Backend}

On the backend, DataDock utilizes the Django web framework. Django is a high-level Python framework that encourages rapid development and clean, pragmatic design \cite{django}. It is known for its robustness, scalability, and extensive community support, making it an ideal choice for building the backend of DataDock.

To facilitate communication between the frontend and backend, DataDock employs the Django REST Framework (DRF). DRF is a powerful and flexible toolkit for building Web APIs \cite{djangorest}. A REST (Representational State Transfer) API is an architectural style for designing networked applications. It defines a set of constraints and properties based on HTTP, such as statelessness, client-server separation, and a uniform interface. RESTful APIs use HTTP methods (GET, POST, PUT, DELETE) to perform CRUD (Create, Read, Update, Delete) operations on resources, which are identified by URLs.

DRF allows for easy creation of API endpoints that can be accessed from various platforms and devices. It provides a set of tools and abstractions for building RESTful APIs quickly and efficiently. DRF's serialization system enables seamless conversion of Django models to various content types, such as JSON or XML, making it ideal for building APIs that can be consumed by different clients.

One of the key benefits of using a RESTful API is that it provides a clear and consistent interface for interacting with the backend. It allows for a separation of concerns between the client and the server, enabling independent development and scaling of the frontend and backend. RESTful APIs also promote statelessness, meaning that each request contains all the necessary information to be processed, making the application more scalable and easier to cache.

For authentication, DataDock uses Knox tokens \cite{knox}, which provide enhanced security compared to the default TokenAuthentication in DRF. Knox tokens are hashed using SHA-512 and encrypted in the database, making them more secure in case of a database compromise. Additionally, Knox tokens have a configurable expiry time, adding an extra layer of security. Token-based authentication is stateless, meaning that the server does not need to keep a record of which users are logged in. This makes the authentication process more scalable and reduces the load on the server.

DataDock utilizes a SQLite database by default, but this can be easily modified to use other popular databases such as MongoDB \cite{mongodb}, PostgreSQL \cite{postgresql}, or MySQL \cite{mysql}. Django's ORM (Object-Relational Mapping) allows for easy interaction with the database through the use of models. A model is a Python class that subclasses django.db.models.Model and represents a single database table. Each attribute of the model represents a database field. This abstraction provides a convenient and intuitive way to define the structure and behavior of the data being stored, while Django handles the underlying database operations.

\subsection{Frontend}
For the frontend, DataDock uses React, a popular JavaScript library for building user interfaces \cite{react}. React's component-based architecture allows for the creation of reusable UI components, which can be easily composed to build complex user interfaces. This modular approach makes the codebase more maintainable and easier to understand. React's virtual DOM (Document Object Model) implementation is another key feature that contributes to its efficiency. Instead of directly manipulating the browser's DOM, React maintains a lightweight copy of the DOM in memory called the virtual DOM. When data changes, React updates the virtual DOM first, calculates the differences (diff) between the virtual and real DOM, and then efficiently updates the browser's DOM with minimal changes. This approach significantly improves performance, especially in large-scale applications with frequent data updates.

React's declarative syntax, JSX (JavaScript XML), allows developers to write HTML-like code within JavaScript. This makes the code more readable and intuitive, as the structure of the UI closely resembles the final rendered output. React's one-way data flow, also known as unidirectional data flow, ensures that data flows in a single direction from parent components to child components. This makes the application's state more predictable and easier to debug.

To manage the application state, DataDock incorporates Redux, a predictable state container for JavaScript apps \cite{redux}. Redux follows the principles of a single source of truth, state immutability, and pure functions for state modifications. It maintains a single, immutable state tree (store) that represents the entire application state. Components can access the required parts of the state by subscribing to the store. To modify the state, components dispatch actions, which are plain JavaScript objects describing the desired changes. Reducers, pure functions that take the current state and an action as input and return a new state, handle these actions and update the state accordingly. This centralized state management makes it easier to reason about the application's behavior, enables features like undo/redo, and facilitates state persistence.

The combination of React and Redux allows for a highly responsive and interactive user interface. When a user interacts with the application, such as creating a new post or rating a dataset, React components dispatch the corresponding actions. Redux processes these actions, updates the state, and notifies the subscribed components. React then efficiently re-renders the affected components, updating the UI in real-time without requiring a full page refresh. This provides a smooth and engaging user experience, similar to that of modern social media platforms.

To communicate with the backend API, DataDock utilizes the Axios library \cite{axios}. Axios is a popular, promise-based HTTP client for JavaScript that simplifies making API requests. It provides an easy-to-use API for sending GET, POST, PUT, and DELETE requests to the server and handles the response data efficiently. Axios also supports request and response interceptors, which allow for easy modification or examination of requests and responses before they are sent or received. This is particularly useful for adding authentication headers, handling errors, or transforming data.

When a user interacts with the application, React components dispatch actions that trigger API requests using Axios. Axios sends the requests to the appropriate backend endpoints, along with any necessary data or parameters. The backend processes these requests, interacts with the database, and returns the response data in JSON format. Axios then resolves the promise, and the received data is dispatched to the Redux store. The store updates the state, and the subscribed React components re-render with the new data, updating the UI accordingly. This seamless integration between React, Redux, and Axios enables a smooth and efficient data flow between the frontend and backend, resulting in a responsive and dynamic user experience.

\subsection{Integration}
The Django backend and React frontend are integrated through REST API calls, enabling seamless communication and data exchange between the two components. When a user interacts with the frontend, such as submitting a form or clicking a button, React components dispatch actions that trigger API requests to the Django backend. These actions are typically handled by Redux middleware, such as Redux Thunk \cite{reduxthunk} or Redux Saga \cite{reduxsaga}, which allows for asynchronous operations and side effects.

The API requests are sent using the Axios library, which is configured with the appropriate backend API endpoints. Axios sends HTTP requests (GET, POST, PUT, DELETE) to the corresponding Django views, along with any necessary data or parameters. The Django views, defined in the views.py files, handle these requests and perform the required processing. They interact with the database using Django's ORM (Object-Relational Mapping) and execute any necessary business logic.

After processing the request, the Django views return the appropriate JSON responses using Django REST Framework's serializers. Serializers convert Django model instances or querysets into JSON format, which can be easily consumed by the frontend. The JSON responses typically include the requested data, such as a list of objects, a single object, or a success/error status.

Axios receives the JSON response from the backend and resolves the promise. The received data is then dispatched to the Redux store using the appropriate action creators. The Redux reducers handle these actions and update the application state accordingly. React components that are subscribed to the relevant parts of the state are automatically re-rendered with the new data, updating the UI in real-time.

This architecture allows for a clear separation of concerns between the frontend and backend. The backend focuses on data processing, business logic, and database management, while the frontend handles user interactions and UI rendering. This decoupling makes the application more maintainable and allows for independent development and scaling of the frontend and backend.

The use of REST APIs provides a standardized and platform-agnostic interface for communication between the frontend and backend. This allows for flexibility in terms of the technologies used on either end. For example, the frontend could be implemented using a different framework or library, such as Angular \cite{angular} or Vue.js \cite{vue}, while still being able to communicate with the Django backend through the same API endpoints. Similarly, the backend could be swapped with a different framework or language, such as Flask \cite{flask} or Node.js \cite{nodejs}, as long as it adheres to the same API contract.

By leveraging the power of Django, Django REST Framework, React, Redux, and Axios, DataDock provides a robust, scalable, and user-friendly platform for researchers and data scientists to manage, share, and collaborate on their data. The integration of these technologies enables a seamless and efficient data flow between the frontend and backend, resulting in a responsive and interactive user experience. The modular architecture and clear separation of concerns facilitate maintainability, extensibility, and independent scaling of the application components.
\section{Current Product}
The current implementation of DataDock provides a fully functional REST API with a range of features tailored for data sharing and collaboration among researchers. Key capabilities of the platform include:
\begin{figure}[h]
  \centering
  \includegraphics[width=\columnwidth]{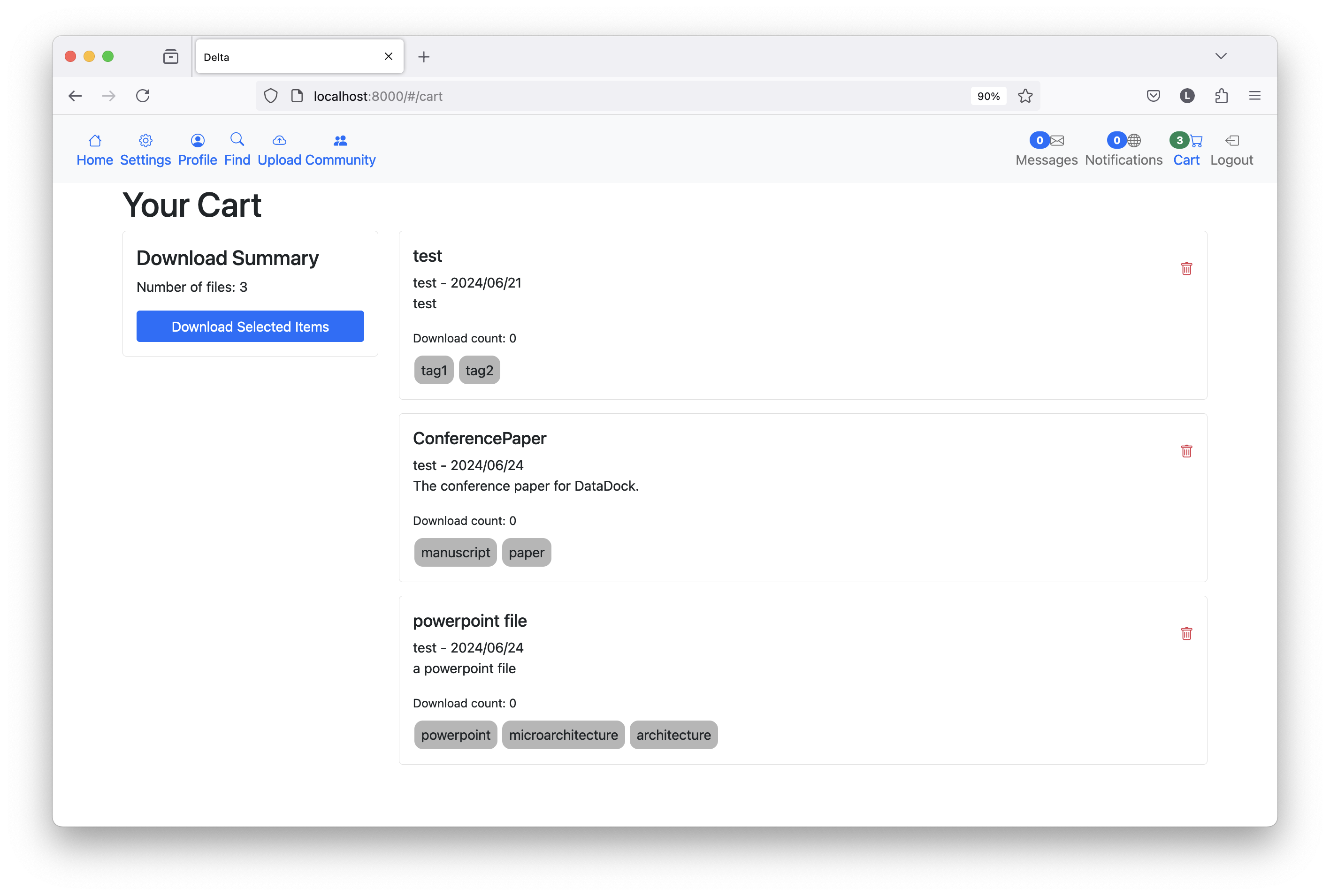}
  \caption{Downloading data. The download page is designed to be similar to that of an e-commerce website.}
  \label{fig:delta-download}
\end{figure}

\subsection{Data Upload and Management}
DataDock enables users to upload datasets through a simple and intuitive interface. The upload process supports both individual files and entire folders, allowing researchers to organize their data hierarchically. When uploading a dataset, users can provide essential metadata such as the dataset name, tags for easy discoverability, and a description to provide context and additional information. Furthermore, users can specify the privacy level of their datasets, choosing from options like public (accessible to all users), public to community (visible only to members of specific communities), or private (accessible only to the dataset owner).
\begin{figure}[h]
  \centering
  \includegraphics[width=\columnwidth]{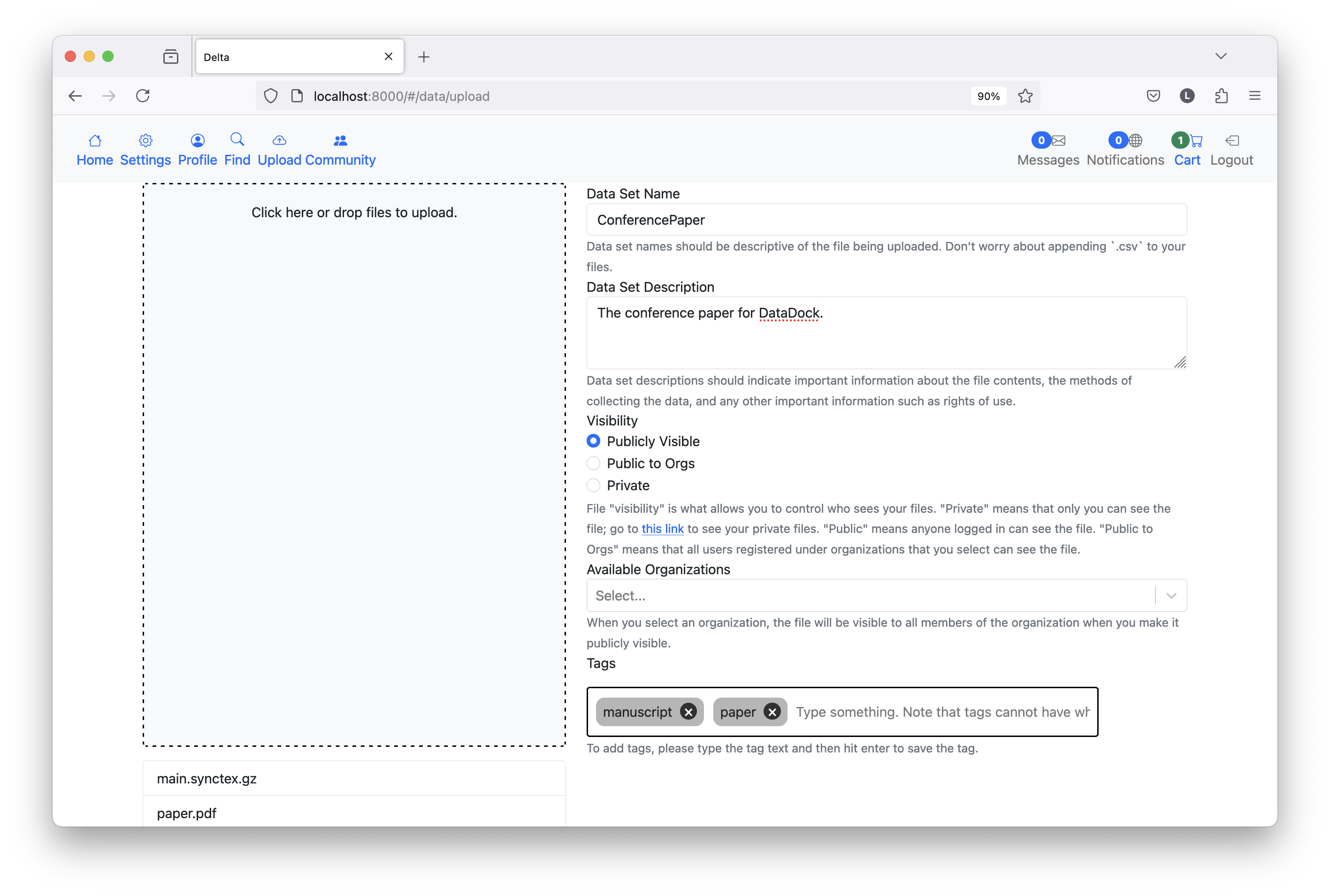}
  \caption{Uploading data. You are able to add a name, description, visibility settings, organization settings, and tags.}
  \label{fig:delta-upload}
\end{figure}

\subsection{User Authentication and Account Management}
The platform incorporates a secure user authentication system, requiring users to register and log in to access the full range of features. User registration is straightforward, and the login process is designed to be seamless and efficient. DataDock also provides account management functionalities, allowing users to update their profile information or remove their account if needed. This ensures that users have control over their personal data and can maintain the accuracy of their profile within the platform.

\subsection{Data Discoverability and Review}
To facilitate data discovery and quality assessment, DataDock implements a rating system for datasets. Users can review and rate datasets on a scale of 1 to 5, providing valuable feedback to dataset owners and helping other researchers gauge the quality and relevance of the data. The reviews are prominently displayed alongside the dataset information, enabling users to make informed decisions when selecting datasets for their research. This feature promotes transparency and encourages the sharing of high-quality data within the research community.
\begin{figure}[h]
  \centering
  \includegraphics[width=\columnwidth]{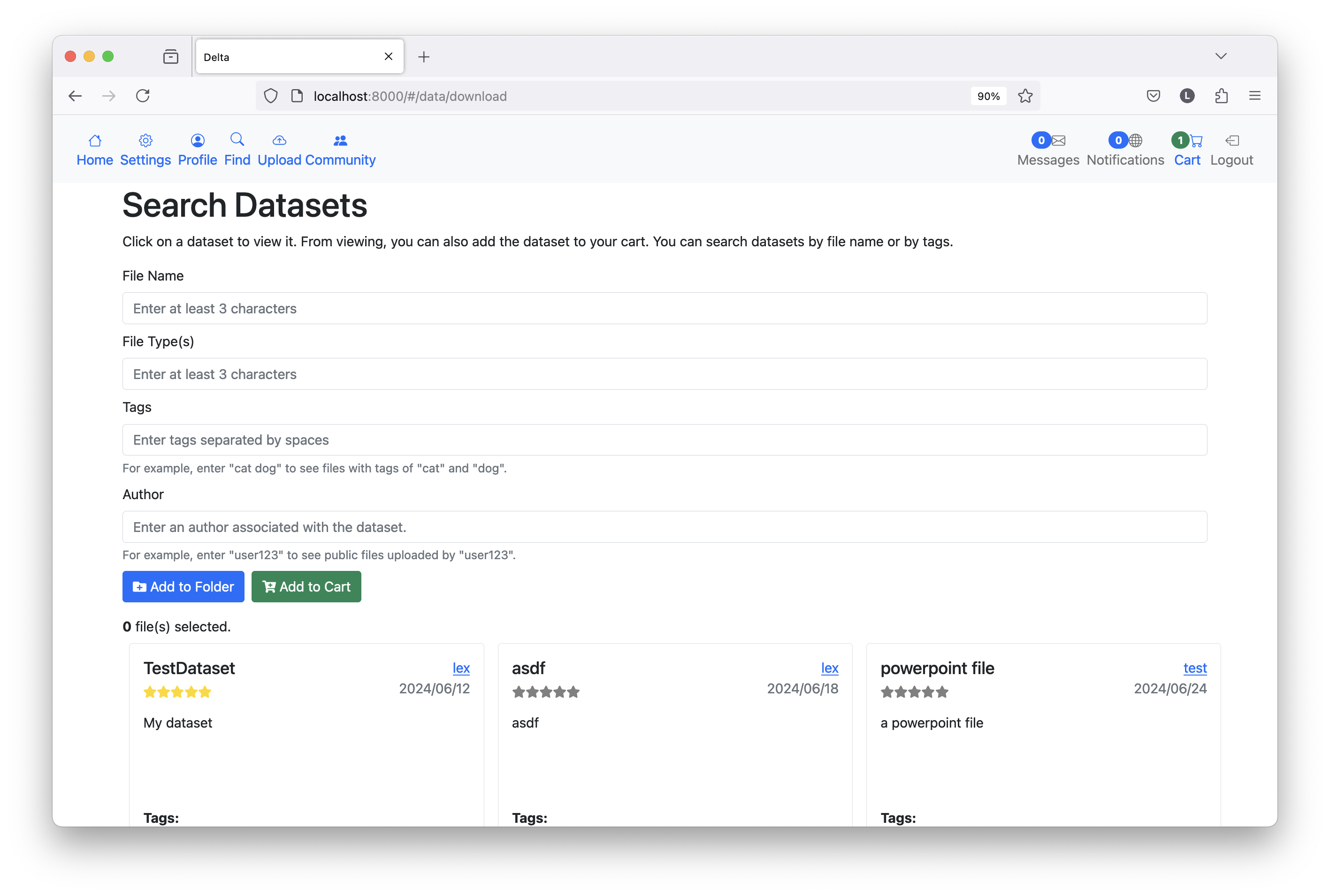}
  \caption{Searching data. You can filter by name, file type, tags, and author.}
  \label{fig:delta-search}
\end{figure}
\begin{figure}[h]
  \centering
  \includegraphics[width=\columnwidth]{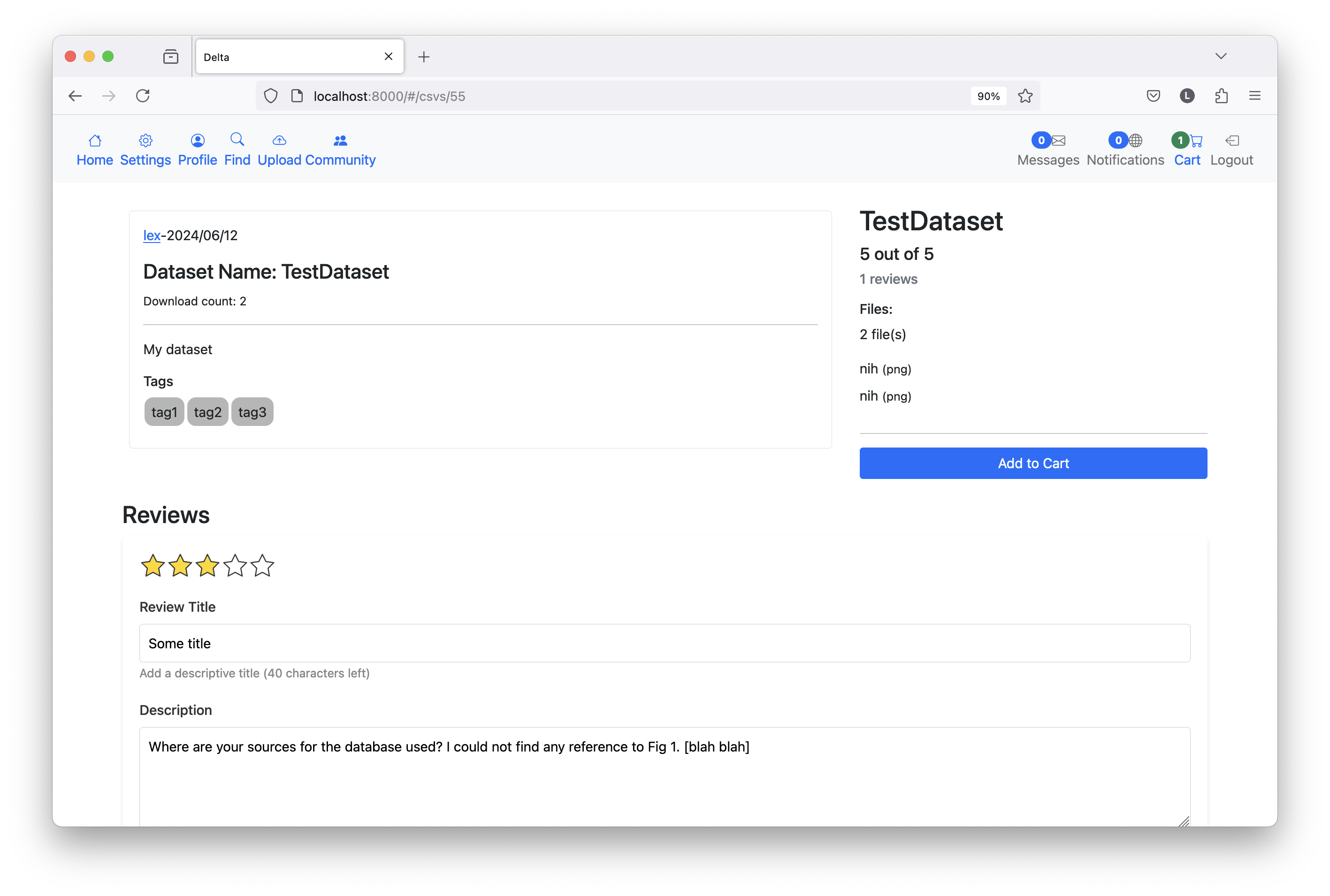}
  \caption{Reviewing data. You can rate the dataset from 1-5, and comment on its utility.}
  \label{fig:delta-review}
\end{figure}

\subsection{Collaboration and Communication}
DataDock recognizes the importance of collaboration and communication in research. The platform offers a direct messaging feature, allowing users to initiate conversations with dataset authors. This facilitates direct interaction between researchers, enabling them to ask questions, provide feedback, or discuss potential collaborations. By fostering communication and knowledge sharing, DataDock aims to break down silos and encourage cross-disciplinary research.
\begin{figure}[h]
  \centering
  \includegraphics[width=\columnwidth]{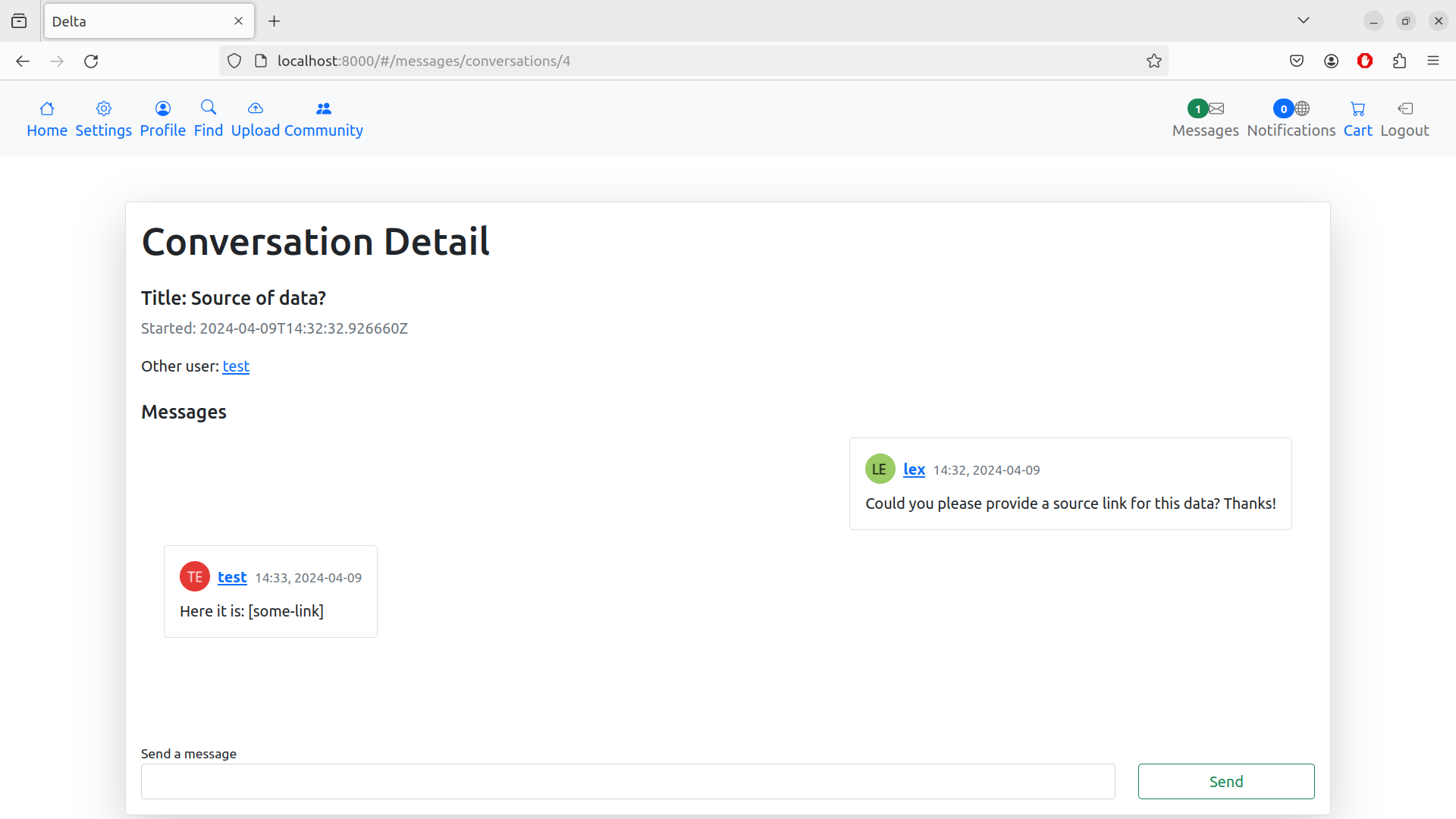}
  \caption{Conversations. You can start a conversation with a user in order to find out more about the data, methods, etc.}
  \label{fig:delta-convo}
\end{figure}

\subsection{Community Building}
To further support collaboration and knowledge exchange, DataDock incorporates community creation and management features. Users can create communities focused on specific research areas, methodologies, or projects. Researchers can join these communities to connect with like-minded individuals, share datasets, and engage in discussions. The community feature enhances the discoverability of relevant datasets and promotes the formation of research networks. By enabling researchers to join and contribute to communities, DataDock facilitates the growth of vibrant and collaborative research ecosystems.
\begin{figure}[h]
  \centering
  \includegraphics[width=\columnwidth]{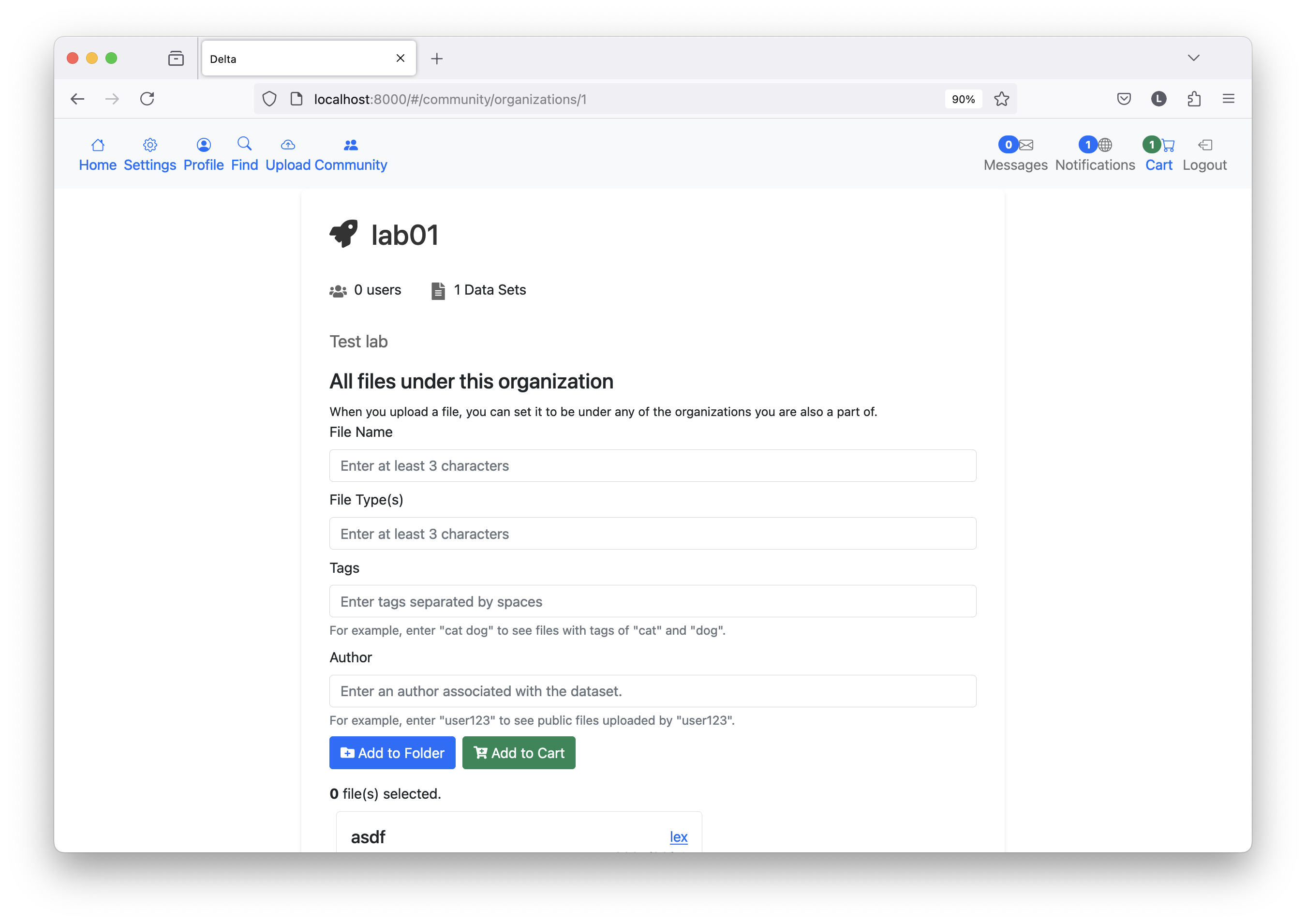}
  \caption{Organizations. When you upload a file as "public to organizations", only organization members would be able to view it.}
  \label{fig:delta-organization}
\end{figure}

\subsection{Notifications and Updates}
DataDock keeps users informed about relevant activities and updates through a notification system. Users receive notifications for various events, such as when a new dataset is uploaded within their communities, when their datasets receive reviews or ratings, or when they receive direct messages from other users. These notifications ensure that researchers stay up to date with the latest developments and can promptly respond to feedback or inquiries. The notification system enhances user engagement and fosters a sense of active participation within the platform.

The current implementation of DataDock demonstrates a comprehensive set of features designed to streamline data sharing, collaboration, and discovery. By providing a user-friendly interface, robust data management capabilities, and communication tools, DataDock empowers researchers to efficiently share their datasets, connect with peers, and advance their research endeavors. The platform's focus on user control, data quality, and community building sets it apart as a valuable tool for the research community.

\section{Summary}
In this paper, we introduced DataDock, an open-source data sharing and collaboration platform designed specifically for research teams. DataDock addresses the limitations of current file storage and sharing services, which often lack features tailored for research data management and can be prohibitively expensive for smaller labs.

DataDock provides a comprehensive solution for storing, organizing, sharing, and collaborating on research data. The platform supports efficient data storage and transfer without incurring extra costs, as it is hosted on a research lab's own server. DataDock's open-source nature allows for customization to suit the unique requirements of any research team and enables the platform to evolve with the needs of the research community.

One of the key advantages of DataDock is its ability to handle heterogeneous data, which is crucial for many research projects. Research labs often work with diverse data types, ranging from structured tabular data to unstructured text, images, and videos. DataDock's agnostic approach to file types ensures that it can accommodate the storage and sharing needs of various research domains, from bioinformatics to particle physics. This versatility makes DataDock an ideal platform for research labs dealing with complex and diverse datasets.

Moreover, DataDock aligns with the goals of funding agencies like the National Science Foundation (NSF), which often support projects involving heterogeneous data. The NSF recognizes the importance of data management and sharing in advancing scientific research and discovery. By providing a robust and flexible platform for managing heterogeneous research data, DataDock can facilitate compliance with funding agency requirements and promote open science practices.

DataDock's user-friendly interface and features like data quality control, collaboration tools, and integration with existing research workflows further enhance its utility for research teams. By streamlining data management and collaboration, DataDock allows researchers to focus on their core research objectives, rather than grappling with the challenges of data storage and sharing.

In conclusion, DataDock fills a critical gap in the research data management landscape by providing an open-source, customizable, and cost-effective solution for storing, sharing, and collaborating on heterogeneous research data. As an adaptable platform that caters to the unique needs of research teams and aligns with funding agency priorities, DataDock has the potential to revolutionize data management practices in the research community and accelerate scientific discovery across disciplines

\bibliographystyle{ieeetr}
\bibliography{refs.bib}

\end{document}